\documentclass[12pt]{article}
\usepackage{fleqn}
\usepackage[dvips]{graphicx}
\usepackage{graphics}
\title
{\bf 
Present Status of the Long Range Component 
 of the Nuclear Force\footnote{NUP-A-2000-10}
}
\vspace{20mm}
\author{Tetsuo Sawada \\
{\small \em Atomic Energy Research Institute, Nihon University, 
Tokyo, Japan 1010062}
\thanks{Associate member of AERI for research. \ \ \  
  e-mail address: t-sawada@fureai.or.jp   }}
\date{}
\begin{document}

\maketitle

\thispagestyle{empty}

\vspace{60mm}
\begin{flushleft}
{\large\bf Abstract}
\end{flushleft}

   In order to settle the fundamental question whether the nuclear 
forces involve the long range components, the S-wave amplitude of the 
proton-proton scattering is analysed in search for the extra 
singularity at $\nu=0$, which corresponds to the long range force. 
To facilitate the search, a function, which is free from the 
singularities in the neighborhood of $\nu=0$ when all the 
interactions are short range, is constructed.  The calculation 
of such a function from the phase shift data reveals a sharp cusp 
at $\nu=0$ in contradiction to the meson theory of the nuclear force. 
The type of the extra singularity at $\nu=0$ is close to what is 
expected in the case of the strong van der Waals interaction.  
Physical meanings of the long range force in the nuclear force are 
discussed.     
 Low energy p-p experiments to confirm directly the strong long 
range interaction are also proposed, in which the characteristic 
 interference patten of the Coulonb and the Van der Waals forces 
 is predicted.   

\newpage

\addtocounter{page}{-1}

\section{Introduction }
     In the Yukawa model of the nuclear force\cite{Yukawa}, the 
pi-meson plays the central roll, since the nuclear potentials are 
assumed to arise from the  exchanges of a pion and a set of pions 
at least outside of the inner core region.    Although the two-pion 
exchange potentials constructed directly from the meson theory are 
not successful in understanding the details of the data of the 
nucleon-nucleon scatterings, various phenomenological nuclear 
potentials have been proposed to reproduce the existing data. 
In the constructions of the nuclear potentials, two features are 
commonly shared by all the nuclear potentials, which are inherited 
from the meson theory of the nuclear force.  One is the one-pion 
exchange (OPE) part of the potential and the other is that the range 
of the non-OPE part is shorter than $1/(2\mu)$.   In terms of the 
singularities of the scattering amplitudes $A(s,t)$, such features 
correspond to the one-pion exchange pole at $t=\mu^2$ and the cut 
of the two-pion exchange which starts at $t=4 \mu^2$.  These analytic 
structures are common to all the amplitudes calculated from the 
various proposed nuclear potentials.
        
    On the other hand, in the nineteen sixties our views on the 
nucleons changed from `elementary particles' to composite 
particles, and moreover in the most important models of hadron, 
such as the QCD and the dyon model\cite{dyon}, the constructive 
forces of the composite particles are the Coulombic types.  
Therefore we cannot exclude  the possibility for the induced long 
range force to appear between the composite particles.   
In particular, in the dyon model of hadron, the appearance of the 
strong van der Waals interaction between hadrons is a natural 
cosequence, because in such a model the hadrons are regarded as 
the `magnetic atoms'\cite{matom}.  In general the long range force 
gives rise to a singularity at $t=0$ in the scattering amplitude 
$A(s,t)$, and therefore the analytic structure is completely 
different from the case of the purely short range interactions. 

    Since there is no apriori reason to believe that the strong \ 
interaction is a synonym for the short range interaction, it is 
desirable to search for the possible long range components of the 
nuclear force, whenever new precise data in $|t| \ll 4\mu^2$ come to 
be available.   It is the difference of the analytic structure that 
allows us to do the search in the clear-cut way without being 
disturbed by the ambiguities such as different parametrization of 
the nuclear potentials in the inner region.   Therefore, our aim in 
the present article is to search for the possible extra singularity 
of $A(s,t)$ at $t=0$, which is characteristic of the long range force,
 or to find the corresponding  extra singularity in the partial wave 
amplitude.   It is important that the search of the extra singularity 
can be carried out independently of  the uncertainties of the spectra 
related to other singularities.   The reason of such possibility 
comes from the fact that the location of the extra singularity is 
the end point of the physical region $ -4 \nu \leq t \leq 0$, where 
the experimental data are available.    Therefore, in contrast to 
the search for the short range force, our search of the extra 
singularity at $t=0$ does not require any analytic continuation from 
the physical region.  In our approach based on the analytic structure, 
the dispersion technique will turn out to be useful.

      In the search of the long range force, the required accuracy 
of the input data depends heavily on the power $\gamma$ of the 
threshold behavior of the spectral function $A_{t}(s,t)$ at $t=0$, 
where for small $\nu$ and $|t|$ the power $\gamma$ is introduced by 
$A_{t}(s,t) = \pi C' t^{\gamma} + \cdots$.    Since $\gamma$ and 
another power $\alpha$, which appears in the asymptotic behavior of 
the long range potential as $V(r) \sim -C/r^{\alpha} $, are related 
by $\alpha=2 \gamma +3$, our observation of the extra singularity 
in the amplitude becomes more and more difficult  and requires the 
higher precision of the input data, as $\alpha$ increases.   
For example, the van der Waals potentials of the London type 
($\alpha=6$) and the Casimir-Polder type ($\alpha=7$) imply 
$\gamma=1.5$ and $\gamma=2.0$ respectively, and therefore to observe 
these van der Waals forces is to recognize the singularities of the 
type  $C'(-t)^{3/2}$ or $C' t^2 \log (-t)$ on the smooth back 
ground function of $A(s,t)$.   In order to observe the long range 
force,  it is essential to obtain the high precision data at least 
in the region of small $|t|$, when the power $\alpha$ of the 
asymptotic potential is not small.

   In the hadron physics, the accuracies of the phase shift data of 
the S-wave proton-proton scattering in the low energy region are 
exceptinally high\cite{Benn}, and so it is better to analyse the 
S-wave amplitude $h_{0}(\nu)$ instead of $A(s,t)$ in our search for 
the long range force. 
 After the partial wave projection, the singularity of $A(s,t)$ 
at $t=t_{1}$ becomes the left hand singularity at $\nu=-t_{1}/4$ in 
$h_{0}(\nu)$.     Therefore the extra singularity at $t=0$ appears 
at $\nu=0$, whereas the OPE pole changes to the logarithmic cut 
starting at $\nu=-\mu^2/4$, and the two-pion exchange spectrum starts 
at $\nu=-\mu^2$ in $h_{0}(\nu)$.   In addition to these 
singularities, $h_{0}(\nu)$ has the unitarity cut in $\nu \geq 0$. 
 Moreover since we are considering the proton-proton scattering, 
 $h_{0}(\nu)$ has also the cuts of the Coulombic interaction  
in $ -\infty <\nu \leq 0$ and of the vacuum polarization in 
$ -\infty < \nu \leq -m_{e}^2$ .

    In section 2, we shall construct a function $K_{0}(\nu)$ whose 
non-OPE part is free from the singularities in $ |\nu| < \mu^2$, when 
the forces are the short range types of the meson theory plus the 
electromagnetic interaction.   If we take advantage that the 
scattering length is known within $0.05\%$, we can even consider 
the once subtracted function 
$K_{0}^{once}(\nu)=( K_{0}(\nu)- K_{0}(0))/\nu$.  
The merit to use the once subtracted function instead of 
$K_{0}(\nu)$ is that, since the power of the threshold behavior at 
$\nu=0$  changes from $\gamma$ to $(\gamma -1)$, the extra 
singularity becomes much easier to be obsreved, when it exists.  
   In section 3, by using the phase shift data we shall evaluate 
numerically the once subtracted Kantor amplitude $K^{once}_{0}(\nu)$, 
and its non-OPE part  $\tilde{K}^{once}_{0}(\nu)$ will be tabulated. 
The result of the evaluation is that $\tilde{K}_{0}^{once}(\nu)$ has 
a sharp cusp at $\nu=0$.

    In sections 4 and 5, $\tilde{K}_{0}^{once}(\nu)$ is fitted by 
using the spectrum of the long range force and that of the short 
range force respectively. It turns out in section 4 that the 
chi-square minimum occurs at $\gamma = 1.48 \sim 1.62$, which is 
close to the $\gamma$ of the van der Waals interaction of the London 
type.  In section 5, it is shown that the conventional spectrum of 
the short range interaction cannot reproduce the cusp. 

    In section 6, by assuming that the well-known mechanism to 
produce the van der Waals force is working, the lower bound of the 
strength $ ^{\ast} e^2$ of the underlying Coulombic force is 
estimated by using the inequality of the strength $C_{6}$ of the 
Van der Waals potential.  The lower bound of $^{\ast}e^2$ becomes 
3.3 or 14 depending on the value of the radius of the composite 
particle, which comes from the measurement of the nucleon form 
factor or the distance where the deviation from the asymptotic 
form of the van der Waals potential becomes appreciable, 
respectively.    In section 6, it is also pointed out that the 
Coulombic interaction between the magnetic monopoles is an imprtant 
candidate of the underlying super-strong Coulombic force, and the 
dyon model of hadron is explained briefly.     Section 7 is used for 
remarks and comments, in which low energy experiments to confirm 
the long range force, by observing the characteristic destructive 
interference pattern of the Coulomb and the strong Van der Waals 
forces, are proposed .               

\section{ Selection of a Regular Function }

     The low energy proton-proton scattering is prominent in its 
accuracy of the measurements in the hadron physics.  Especially the 
S-wave amplitude $h_{0}(\nu)$ in the low energy region provides the 
ideal place to answer to the fundamental question whether the 
hadron-hadron interactions are short range except for the 
electromagnetic components. In the following investigations, we shall 
make use of the difference of the analytic structures of the 
amplitudes $h_{0}(\nu)$.    In general, for the short range 
interaction which arises from the exchange of 
 a state with mass $\sqrt{t_{1}}$, a singularity appears at 
 $\nu=-t_{1}/4$ in the partial wave amplitude $h_{0}(\nu)$.  
 On the other hand, the long range interaction gives rise to a 
 singularity at $\nu=0$.  Since the location of the extra singularity 
due to the long range force is the end point of the physical region 
$\nu \geq 0$, where the experimental data are available, we can 
expect to observe the singularity directly, if it exists, without 
making any analytic continuation from the physical region.  
The first thing we have to do is to construct a function which is 
regular at $\nu=0$, when all the interactions are short range.  
Next thing is to eliminate the known near-by  singularities from 
the function.   The function with such a wider domain of analyticity 
serves to expose the extra singularity at $\nu=0$, 
and makes it easier for us to observe the long range interaction, 
when it exists.

      Although the main aim of this section is to construct such 
an analytic function for the proton-proton scattering where the 
effects of the vaccum polarization as well as the Coulomb interaction 
are not negligible, it is instructive to start by constructing 
such a function for the neutron-neutron scattering first.  
It is well-known that the S-wave amplitude $h_{0}(\nu)$ has the 
unitarity cut in $\nu > 0$ with the spectral function  
Im$h_{0}(\nu)$, where $h_{0}(\nu)$ relates to the phase shift by
\begin{equation}
  h_{0}(\nu)=\frac{\sqrt{m^2+\nu}}{\sqrt{\nu}} 
  e^{i \delta_{0}(\nu)} \sin \delta_{0}(\nu) \quad .
\end{equation}  
The most famous function, which is analytic at $\nu=0$ and therefore 
accepts the Taylor expansion of $\nu$ when all forces are short 
range, is the effective range function $X_{0}(\nu)$ of Bethe.   
As it is well-known, $X_{0}(\nu)$ is defined by
\begin{equation}   
  X_{0}(\nu)= \sqrt{\nu} \cot \delta_{0}(\nu) .
\end{equation}
From Eqs.(1) and (2), the relation between $h_{0}(\nu)$ and 
$X_{0}(\nu)$ is 
\begin{equation}   
 h_{0}(\nu)= \frac {\sqrt{m^2+\nu}}{ X_{0}(\nu) -i \sqrt{\nu}} \quad .
\end{equation}
 
      Since the one-pion exchange (OPE) contribution is
\begin{equation}  
 h_{0}^{1\pi}(\nu)=\frac{1}{4} \frac{g^2}{4 \pi} \frac{\mu^2}{4 \nu} 
 \log (1 + \frac{4 \nu}{\mu^2}) \quad ,
\end{equation}
the values of the coupling constant $g^2/4 \pi$ and the neutral 
pion mass $\mu$ are sufficient to eliminate the OPE cut from 
$h_{0}(\nu)$.   In fact $\tilde{h}_{0}(\nu)\equiv h_{0}(\nu)-
h_{0}^{1 \pi}(\nu)$ does not have the OPE cut. 
 However Eq.(3) indicates that in order to eliminate the OPE cut 
from $X_{0}(\nu)$ information on Re$ \, h_{0}(\nu)$ in 
$\nu \leq -\mu^2/4$ as well as $g^2/4 \pi$ and $\mu$ are necessary.   
Therefore the effective range function $X_{0}(\nu)$ is not 
adequate for our purpose to construct a function with wider 
domain of analyticity. 

    Another possibility is the Kantor amplitude\cite{Kantor}, 
 which is defined by
\begin{equation}    
K_{0}(\nu) = h_{0}(\nu) - \frac{1}{\pi} \int_{0}^{\infty} d \nu' 
\frac{{\rm Im} \; h_{0}(\nu')}{\nu' -\nu} \quad .
\end{equation}
 The Kantor amplitude does not have the unitarity cut, and the 
values of  $K_{0}(\nu)$ for real positive $\nu$ can be evaluated 
directly from the experimental data, although the integration must 
be regarded as the principal value integration of Cauchy. 
 It is straightforward to eliminate the OPE cut from the Kantor 
amplitude, and which is achieved by introducing the non-OPE 
Kantor amplitude $\tilde{K}_{0}(\nu)$ by
 \begin{equation} 
 \tilde{K}_{0}(\nu)= K_{0}(\nu) - K_{0}^{1 \pi}(\nu) \quad ,
 \end{equation}
where the OPE part of the Kantor amplitude is common to that 
of the partial wave amplitude, namely $K_{0}^{1 \pi}(\nu)= 
h_{0}^{1 \pi}(\nu)$.
It is useful to rewrite the Kantor amplitude introduced in 
Eq.(5) into the form of the contour integration,
\begin{equation}  
K_{0}(\nu)= - \frac{1}{2 \pi i} \int_{C}  d \nu' 
\frac{ h_{0}(\nu')}{\nu' -\nu} \quad ,
\end{equation}
where the closed contour $C$ is shown in Figure 1.  
The merit to write $K_{0}(\nu)$ in the form of Eq.(7) is that 
because of Eq.(3), when the effective range function $X_{0}(\nu)$ 
is a polynomial of $\nu$ or more generally a meromorphic function 
of $\sqrt{\nu}$, then the only singularities of $h_{0}(\nu)$ on 
the first sheet of $\nu$ are poles.\footnote{Strictly speaking 
$\sqrt{m^2+\nu}$ of the numerator of Eq.(3) causes a branch point 
at $\nu=-m^2$.

  Since $\nu=-m^2$ is the far-away singularity, it does not affect 
our search for the extra singularity at $\nu=0$.  However if we want 
to be more precise, we may redefine the amplitude by multiplying 
$m/\sqrt{m^2+\nu}$ to $h_{0}(\nu)$, and make necessary changes in 
the following calculations.}   
Therefore by shrinking the contour $C$ to a point, $K_{0}(\nu)$ 
becomes the sum of the contributions of the poles on the first sheet 
of $\nu$ or in the upper half $\sqrt{\nu}$-plane.    In this way, 
we need not carry out the principal value integration of the 
rapidly changing function of Eq.(5).
 This technique to change the integration into the sum of the 
 contributions of poles will be used in the actual calculation 
 of $K_{0}(\nu)$ in the next section. We are now in the position 
 to examine $\tilde{K}_{0}(\nu)$ in search for the extra singularity 
 at $\nu=0$, if the precise data of the neutron-neutron scatterings 
 were available.  
     
 \begin{figure}[htbp]
 \includegraphics[width=11.5cm,height=7.0cm]{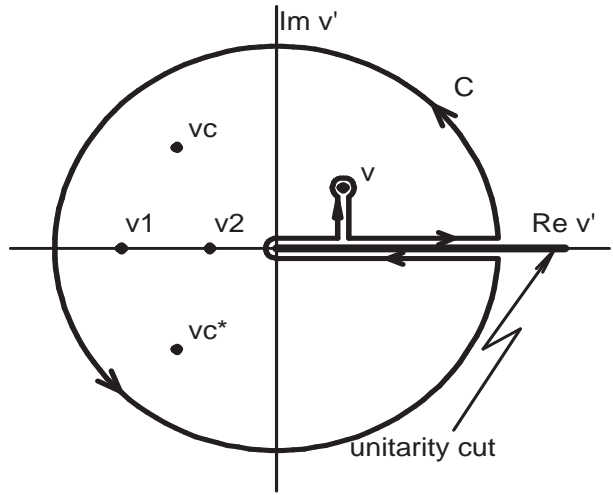}
\caption{{\footnotesize
  The contour $C$ of the integration in Eq.(7).   $\nu_{1}$ 
  and $\nu_{2}$ are real poles on negative real axis, whereas 
  $\nu_{c}$ and  $\nu_{c}^{\ast}$ are a pair of complex poles 
  on the $\nu'$-plane. Their values are listed in Table 2. }}
\end{figure}

     Let us turn to the proton-proton scattering, where the 
vacuum polarization as well as the Coulombic interactions are 
important.  The Kantor amplitude introduced in Eq.(5) does not 
satisfy our requirement of the analyticity.   This is because the 
Coulombic cut in $ -\infty < \nu \leq 0 $ and the cut of the vaccum 
polarization in $-\infty < \nu \leq -m_{e}^2$ still remain in the 
Kantor amplitude $K_{0}(\nu)$ of Eq.(5).   The difficulties are 
by-passed if we remember the modified effective range function 
$X_{0}(\nu)$ of the proton-proton scattering, which is regular at 
$\nu=0$ and accepts the effective range expansion, when all the 
forces are short range except for the terms of the Coulomb and of 
the vacuum polarization.   The modified effective range function 
$X_{0}(\nu)$ for the phase shift $\delta_{0}^{E}(\nu)$ is
\begin{equation}  
 X_{0}(\nu)= \frac{C_{0}^2 \sqrt{\nu}}{1-\phi_{0}} \{ (1+\chi_{0}) 
 \cot \delta_{0}^{E} -\tan \tau_{0} \} +m e^2 h(\eta) +m e^2 
 \ell_{0}(\eta) \quad .
\end{equation}
 In Eq.(8), two well-known functions with the Coulombic order of 
 magnitudes appear, they are expressed using a new variable 
 $\eta=m e^2/(2 \sqrt{\nu})$ :
\begin{equation}   
C_{0}^2 = \frac{2 \pi \eta}{ e^{2 \pi \eta}-1} \qquad \; and \qquad 
\; h(\eta)= \eta^2 \sum_{\ell =1}^{\infty} \frac{1}
{\ell (\ell^2+ \eta ^2)} -\log \eta - 0.57722 \cdots .
\end{equation}
In Eq.(7) $\tau_{0}$ is the phase shift due to the vacuum 
polarization potential\cite{vacpol}
\begin{equation}  
 V^{vac}(r)= \lambda \frac{e^2}{r} \int_{4 m_{e}^2}^{\infty} dt 
 \frac{ e^{-r \sqrt{t}}}{2 t} (1+ \frac{2 m_{e}^2}{t}) 
 \sqrt{1-\frac{4 m_{e}^2}{t}}
\equiv \lambda  \frac{e^2}{r} I(r) \quad ,
\end{equation}
where  $m_{e}$ is the mass of the electron and $\lambda=2 e^2/3 
\pi =1.549 \times 10^{-3}$.  Functions $\tau_{0}$, $\chi_{0}$, 
$\phi_{0}$ and $\ell_{0}(\eta)$ have the order of magnitudes of the 
vaccum polarization, and  it is sufficient for our purpose to retain 
only the first term in $\lambda$.   In such approximation, they are 
expressed by $F_{0}(r)$ and $G_{0}(r)$, which are the regular and the 
irregular Coulombic wave functions respectively, as  
\begin{equation}   
  \tau_{0} = -2 \eta \lambda \int^{\infty}_{0} dr \frac{F_{0}(r)^2 
  I(r)}{r} \\
\end{equation}
\begin{equation}  
  \chi_{0} =  \phi_{0}=-2 \eta \lambda \int_{0}^{\infty} dr 
  \frac{F_{0}(r) G_{0}(r) I(r)}{r} \\
\end{equation}
\begin{equation}  
 \ell_{0} (\eta) = -\lambda \int_{0}^{\infty} dr \frac{I(r)}{r}
  [ (C G_{0}(r))^2- (C G_{0}(r))^{2}_{\nu=0}] \quad ,
 \end{equation}
where $I(r)$ is defined in Eq.(10).   Exact definitions of these 
functions are found in the paper by Heller\cite{vacpol}.

      By using the modified effective range function $X_{0}(\nu)$, 
we define the S-wave amplitude $h_{0}(\nu)$ of the p-p scattering by
\begin{equation}  
h_{0}(\nu)=\frac{\sqrt{m^2+\nu}}{ X_{0}(\nu) - m e^2 h(\eta) -i 
\sqrt{\nu} C_{0}^2} \quad .
\end{equation}
 The relation between $h_{0}(\nu)$ and the phase shift 
 $\delta_{0}^{E}$ is obtained if we substitute $X_{0}(\nu)$ of 
 Eq.(8) into Eq.(14), and which reduces to the well-known form   
\begin{equation}  
h_{0}(\nu)=\frac{1}{C_{0}^2} \frac{\sqrt{m^2+\nu}}{\sqrt{\nu}} 
e^{i \delta_{0}^{E}(\nu)} \sin \delta_{0}^{E}(\nu) \quad ,
\end{equation}
if the functions related to the vacuum polarization are neglected.  
 The form of $h_{0}(\nu)$ of Eq.(15) is the same as that of the 
 neutron-neutron scattering of Eq.(1) except for the factor 
 $C_{0}^2$ given in Eq.(9), which is the penetration 
 factor.   
If we compare the S-wave amplitude of the p-p scattering with 
that of the n-n scattering, which are given in Eq.(14) and Eq.(3) 
respectively, a combination of functions $(-m e^2 h(\eta) - 
i \sqrt{\nu} C_{0}^2)$ appears in place of $\ -i  \sqrt{\nu}$.  
In order to investigate the analytic structure of $h_{0}(\nu)$, 
it is convenient to rewrite the combination as
\begin{equation}   
 -m e^2 h(\eta) - i \sqrt{\nu} C_{0}^2 = -i \sqrt{\nu} + m e^2 
 \{\log (i \eta) -\psi(1+i \eta) \} \quad .
\end{equation}
Since the digamma function $\psi(z)$ has poles at non-positive 
integers, the poles on the $\eta$-plane appear on the positive 
imaginary axis. 
In terms of $\sqrt{\nu}$, which is $m e^2/(2 \eta)$, the series of 
poles appear on the negative imaginary axisis and converge to 
$\sqrt{\nu}=0$. It is the smallness of the fine structure constant 
$e^2$ and therefore of the residues of such poles that zeros of the 
denominator of Eq.(14) occur at points very close to the locations 
of the poles of $ (-m e^2 h(\eta)-i \sqrt{\nu}C_{0}^2 )$.  
Therefore the partial wave amplitude $h_{0}(\nu)$  of the p-p 
scattering has a series of poles on the second sheet of $\nu$, 
namely on the lower half plane of $\sqrt{\nu}$, whereas on the first 
sheet of $\nu$ the analytic structure of $h_{0}(\nu)$ does not change 
compared to the case of the n-n scattering.
  This fact implies that the same definition of the Kantor amplitude 
  $K_{0}(\nu)$ introduced for the neutron-neutron scattering, which 
  is given in Eq.(5), is valid also for the proton-proton scattering, 
 as long as we evaluate Im$h_{0}(\nu')$ of Eq.(5) from Eqs.(14) and 
  (16).
  Therefore the Kantor amplitude of the p-p scattering $K_{0}(\nu)$ 
  constructed in this way is free from the singularities in the 
  neighborhood of $\nu=0$, and so does not have the cut of the 
  vacuum polarization as well as that of the Coulomb interaction.

 \section{Numerical Calculation of the Kantor \\
  Amplitude }

     Since the scattering length of the S-wave amplitude of the p-p 
scattering is known with high precision, we shall analyse the once 
subtracted Kantor amplitude, which is
\begin{equation}  
   K_{0}^{once}(\nu) \equiv \frac{K_{0}(\nu) - K_{0}(0)}{\nu}.
\end{equation}
 Merits to use the once subtracted Kantor amplitude rather than 
 $K_{0}(\nu)$ are twofold.  The first one is the higher convergence 
 of the integration  
\begin{equation} 
K_{0}^{once}(\nu) = \frac{h_{0}(\nu)-h_{0}(0)}{\nu}- \frac{1}{\pi} 
\int_{0}^{\infty} d \nu' \frac{ Im \; h_{0}(\nu')}{ \nu' (\nu' - \nu)}
\quad ,
\end{equation}
namely the uncertainty arising from the lack of the very accurate 
data in the higher energy region is largely suppressed because of the 
extra factor $\nu'$ in the denominator of the integrand.  The second 
one is the change of the power of the threshold behavior of the 
spectral function from $C'(-\nu')^{\gamma}$ to $- C' 
(-\nu')^{\gamma -1}$, and which makes it easier for us to recognize 
the extra singularity at $\nu=0$ in $K_{0}^{once}(\nu)$.  
Since the power $\alpha$, which appears in the asymptotic behavior 
of the potential as  $V(r) \sim -C_{\alpha}/r^{\alpha}$, and the 
power $\gamma$ of the spectral function Im$ \: h_{0} (\nu')$ are 
related by
\begin{equation}  
    \alpha= 2 \gamma +3  \quad ,
\end{equation}
 $\gamma$ is equal to $3/2$ for the van der Waals potential of the 
London type, and which is expected to occur in the dyon model of 
hadron\cite{dyon}\cite{matom}.    Therefore in such a case the 
behavior of the once subtracted Kantor amplitude in the neighborhood 
of $\nu=0$ is that 
\begin{equation} 
   K_{0}^{once}(\nu) = {\rm const.} + C' \sqrt{\nu}+\cdots  
   \qquad {\rm with} \quad C'>0 \quad ,
\end{equation}
and we must observe the extra singularity as a cusp at $\nu=0$ 
unless the coefficient $C'$ is very small.    
    
      Let us represent the effective range function of the p-p 
scattering $X_{0}(\nu)$ by a meromorphic function of $\nu$:
\begin{equation}  
   \frac{X_{0}(\nu)-c_{0}-c_{1} \nu}{\nu} (1- \frac{\nu}{\nu_{P}}) 
= a_{0} + a_{1} \nu + \frac{b_{0} + b_{1} \nu}{d_{0} +d_{1} 
\nu +\nu^2} \quad ,
\end{equation}
in which $\nu_{P}$ is the location of the pole of $X_{0}(\nu)$. 
$-c_{0}$ and $c_{1}$ are the inverse of the scattering length and 
the effective range devided by 2 respectively.   
The r.h.s. of Eq.(21) involves six free parameters to be fitted.   
In the fitting, the Compton wave length of the neutral pion will be 
used as the unit of the length.  In the search of the free 
parameters, $c_{0}$, $c_{1}$ and $\nu_{P}$ are fixed beforehand 
which are\cite{nij88}
\begin{equation}  
c_{0}=0.18698 \quad , \quad c_{1}=0.9541 \quad and \quad
  \nu_{P}=6.39 \qquad .
\end{equation}
 Accuracies of $c_{0}$ and $c_{1}$ are as high as 0.05\% and 0.6\%
 respectively.  On the other hand, uncertainty of $\nu_{P}$ is 
 around 3 per cent.    However reasonable accuracy of $\nu_{P}$ 
 is sufficient for our purpose to search for the extra singularity 
 at $\nu=0$ . 
          
    The S-wave phase shifts of the p-p scattering by the Nijmegen 
group\cite{nij88}\cite{nij90}\cite{nij93} are used as the input 
to evaluate the effective range function $X_{0}(\nu)$.   
The once subtracted form of $X_{0}(\nu)$, in which the pole at 
$\nu=\nu_{P}$ is eliminated by multiplying a factor 
$(1- \nu/\nu_{P})$, is displayed in figure 2.  The points in 
the figure come from the multi-energy phase shifts of 
Nijmegen-90 \cite{nij90}.

 \begin{figure}[htbp]
\begin{minipage}{6.8cm}
\includegraphics[width=.99\textwidth,height=5.0cm]{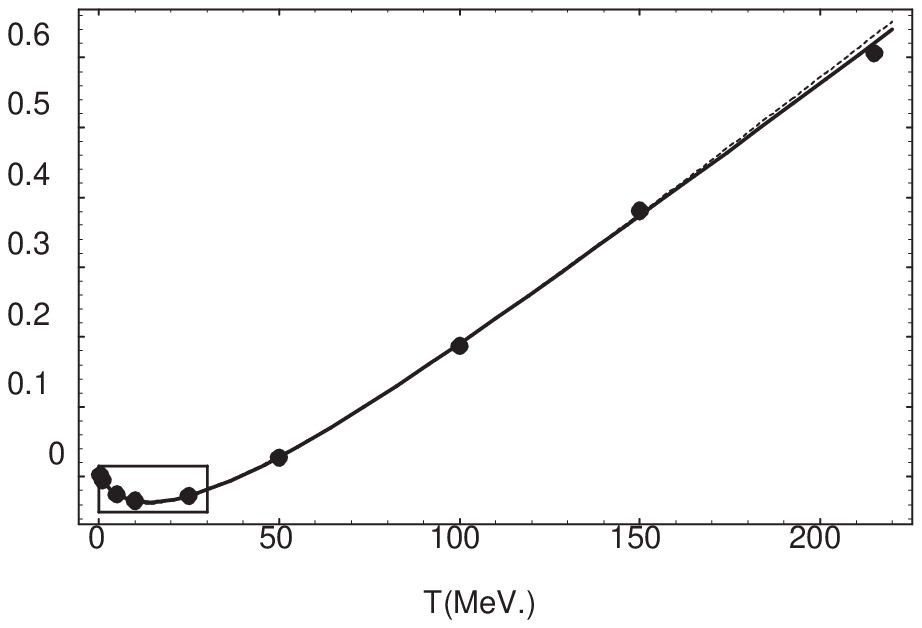}
\caption{{\footnotesize
 $(X_{0}(\nu)-c_{0}-c_{1}\nu) (1-\nu/\nu_{P})/\nu$ of Eq.(21) 
 is plotted against $T_{lab}$, where $X_{0}(\nu)$ is the effective 
 range function  of the S-wave p-p scattering.   
 The points are the data of Nijmegen 90. The box of the left 
 lower corner is enlarged in fig.3.}}
\end{minipage}
\hfill
\begin{minipage}{6.8cm}
\includegraphics[width=.99\textwidth,height=5.0cm]{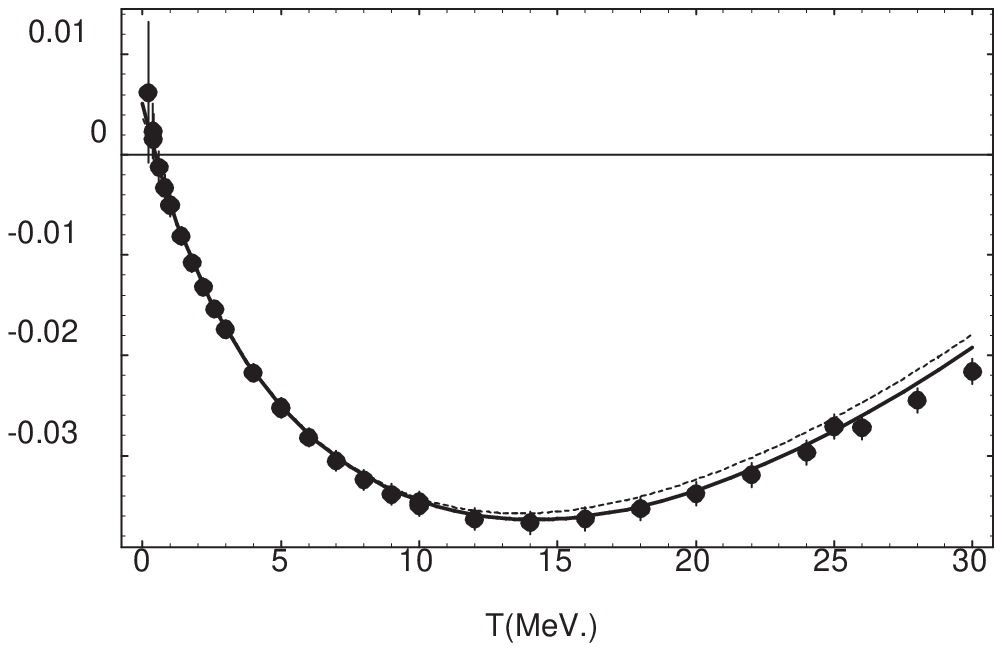}
\caption{{\footnotesize  Enlarged graph of 
$(X_{0}(\nu)-c_{0}-c_{1}\nu) (1-\nu/\nu_{P})/\nu$ with error bar 
is plotted against $T_{lab}$ in $T_{lab} \leq 30 \; MeV.$, 
in which the points are the data of Nijmegen 88 and 90.    
The parameters of the fitted curves are listed in Table 1.}}
\end{minipage}
\end{figure}
 The box at the left lower corner in 
the figure is enlarged and displayed in figure 3, and in which 
the low energy data in the energy range  $0.1 \; MeV. \leq T_{lab} 
\leq 30 \; MeV.$ of Nijmegen-88 \cite{nij88} are shown with their 
error bars.    The data in $0.5 \; MeV. \leq T_{lab} \leq 100 \; 
MeV.$ are fitted by the meromorphic function of Eq.(21), and the 
six parameters are determined by the chi square search.  
         The searched curve is shown in figures 2 and 3 as the full 
line, and whose parameters are tabulated in the column [sc] in 
Table 1.   On the other hand the dash curve in the figures is a 
different fit by a meromorphic function, in which the six parameters 
are fixed by the six data points at $T_{lab}=$ 1, 5, 10, 25, 50, 100 
MeV. of the Nijmegen-90\cite{nij90}.   
The parameters of the fixed curve are tabulated in the column [fc] 
in Table 1.    

\begin{table}[htb]
\begin{center}
\begin{minipage}{9.5cm}  
   \renewcommand{\footnoterule}{}
   \begin{center}
     \caption{Parameters of the meromorphic form of the 
     effective range function introduced in Eq.(21)}
     \vspace{2.5mm}
\begin{tabular}{|c|lll|l|}
\hline
   & \ \ [uc] \ &\ \ [sc] \ &\ \ [$l  $c] \ &\ \ [fc]  
   \rule[-0.3em]{0em}{1.7em} \\
\hline
 $a_{0}$ \  & $-$0.2579 & $-$0.2703 & $-$0.2815 & $-$0.3295 
 \rule{0mm}{1.3em} \\
 $a_{1}$ \  &\ \ 0.15278 &\ \ 0.15504 &\ \ 0.15694 &\ \ 0.16328  \\
 
 $b_{0}$ \  &\ \ 0.01009 &\ \ 0.01983 &\ \ 0.03463 &\ \ 0.09272 
 \rule{0mm}{1.3em}  \\
 $b_{1}$ \  &\ \ 0.1862 &\ \ 0.2098 &\ \ 0.2337 &\ \ 0.39475 \\
 
 $d_{0}$ \  &\ \ 0.0379 &\ \ 0.0720  &\ \ 0.1218 &\ \ 0.2782 
 \rule{0mm}{1.3em} \\
 $d_{1}$ \  &\ \ 0.812 &\ \ 0.927 &\ \ 1.061 &\ \ 1.643 \\
\hline
\end{tabular}    
      \end{center}
    \end{minipage}
  \end{center}     
\end{table}
In the same table, there are also columns of the 
curves of [$l$c] and [uc], in which the parameters are determined 
by fitting to the lower fringe and to the upper fringe 
of the error bars in $0.5 MeV. \leq T_{lab} \leq 100 \; MeV.$ 
respectively.   Such parameters are necessary to estimate 
the errors of the Kantor amplitude.

   It is remarkable that in figures 2 and 3, the curvature of 
 the curve in the small $T_{lab}$ region increases very rapidly 
as $T_{lab}$ decreases to zero.   If we remember that the starting 
points of the spectra of the one-pion exchange(OPE) and of the 
two-pion exchange(TPE) are $T_{lab}= -9.71 \; MeV.$ and $ -38.83 
\; MeV.$ repectively, it is not easy to understand the rapid change 
of the curvature in the low energy region such as $ 0 < T_{lab} <10 
\; MeV.$ .   On the other hand, if we accept the long range force, 
the appearance of such a cusp at $T_{lab}=0$ is what is expected.   
Since the separation of the OPE term from $X_{0}(\nu)$ is not 
simple, in the present article we shall examine 
$K_{0}^{once}(\nu)$ because, contrary to 
$X_{0}(\nu)$, the separation of the OPE term from the Kantor 
amplitude is straightforward. 
 
 \begin{table}[htb]
\begin{center}
\begin{minipage}{10.8cm}  
   \renewcommand{\footnoterule}{}
   \begin{center}
     \caption{Poles and residues of $h_{0}(\nu)$ for meromorphic 
       $X_{0}(\nu)$}
     \vspace{2.5mm}
 \begin{tabular}{|c|lll|l|}
\hline
   &\ \ [uc] \ &\ \ [sc] \ &\ \ [$l $c] \ &\ \ [fc]  
   \rule[-0.3em]{0em}{1.7em} \\
\hline
 $\nu_{1}$ \  & $-$0.04951 & $-$0.08487 & $-$0.12894 & $-$0.18430 
 \rule{0mm}{1.5em} \\
 $r_{1}$ \  &\ \ 0.005345 &\ \ 0.016313 &\ \ 0.044157 &\ \   0.15756 \\
 
 $\nu_{2}$ \  & $-$0.54699  & $-$0.58917 & $-$0.63853  & $-$0.87572 
 \rule{0mm}{1.5em}\\
 $r_{2}$ \  &\ \ 2.5629 &\ \ 2.8719 &\ \ 3.1755 &\ \   5.0604 \\
 
 Re$ \; \nu_{c}$ \  & $-$3.4550 & $-$2.8526 & $-$2.4303 & $-$1.3318 
 \rule{0mm}{1.5em} \\
 Im$ \; \nu_{c}$ \  &$-$6.0285 &$-$6.0191 &$-$5.9984 & $-$6.0552\\
Re$\; r_{c}$ \  & $-$25.427 & $-$22.461 &$-$20.562  & $-$16.855 
\rule{0mm}{1.2em} \\
 Im$\; r_{c}$ \  &\ \ 43.256 &\ \ 36.108 &\ \ 31.607 &\ \ 21.008  \\
\hline
\end{tabular}
       \end{center}
    \end{minipage}
  \end{center}     
\end{table}

   In Table 2, the locations of the poles of $h_{0}(\nu)$, which is 
introduced in Eq.(14), on the first sheet of $\nu$ and their residues 
are  tabulated.  The columns correspond to those in Table 1, 
and each of them has two real poles $ \nu_{1}$ and $\nu_{2}$, 
and a pair of complex poles $(\nu_{c} , \nu_{c}^{\ast})$ 
respectively, on the first sheet of $\nu$.

    In Table 3, $\tilde{K}_{0}^{once}(\nu)$, which are the once 
subtracted Kantor amplitude of Eq.(17) minus the one-pion exchange 
contribution, are tabulated for [sc] and [fc].   Because of Eq.(7),
\begin{equation}  
 \tilde{K}_{0}^{once}(\nu)=\frac{r_{1}}{\nu_{1} (\nu -  \nu_{1})} 
 + \frac{r_{2}}{\nu_{2} (\nu - \nu_{2})}+ 2 Re \{ \frac{r_{c}}
 {\nu_{c} (\nu - \nu_{c})} \} +\Delta_{0} - (OPEC)
\end{equation}
, where OPEC is the one-pion exchange contribution.  
The correction term $\Delta_{0}$ arises from the approximation to 
$X_{0}(\nu)$ by the meromorphic function fitted to the phase shift 
in the energy range of $ 0.5 \;MeV. < T_{lab} < 100 \; MeV.$.   
\vspace{8mm}
\begin{table}[htb]
\begin{center}
\begin{minipage}{15.0cm} 
    \renewcommand{\footnoterule}{}
    \begin{center}
    \caption{ Non-OPE part of the once subtracted Kantor amplitude 
    $\tilde{K}_{0}^{once}(\nu)$ }
\vspace{5mm}

\begin{footnotesize}
\begin{tabular}{|c|c|c|c|c||c|c|c|c|c|}
\hline
$\sqrt{\nu}$&$T_{lab}$&\multicolumn{2}{c|}{$\tilde{K}_{0}^{once}
(\nu)$}& \rule{0em}{1.2em} &
$\sqrt{\nu}$&$T_{lab}$&\multicolumn{2}{c|}{$\tilde{K}_{0}^{once}
(\nu)$}& \rule{0em}{1.2em} \\ \cline{3-4} \cline{8-9}
$(\mu \! = \! 1)$\rule[-0.2em]{0em}{1.2em} &(MeV.)& [$\,sc \,$] & 
[$\, f\, c\,$] &\raisebox{0.4em}{$\Delta \tilde{K}_{0}^{once}$}&
$(\mu \!= \! 1)$\rule[-0.2em]{0em}{1.2em} &(MeV.)& [$\,sc \,$] & 
[$\, f\, c\,$] &\raisebox{0.4em}{$\Delta \tilde{K}_{0}^{once}$}
\\
 \hline 
 \hline

0.05 & 0.097 & -5.188 & -4.974 & 0.2803 \rule{0mm}{1.2em}&
1.15 & 51.35 & -2.651 & -2.655 & 0.0126 \rule{0mm}{1.2em}\\ 
0.10 & 0.388 & -5.049 & -4.892 & 0.2069 &
 1.20 & 55.92 & -2.581 & -2.584 & 0.0116  \\
0.15 & 0.874 & -4.864 & -4.771 & 0.1307  &
1.25 & 60.67 & -2.515 & -2.517 & 0.0105 \\
0.20 & 1.553 & -4.671 & -4.629 & 0.0747 &
1.30 & 65.62 & -2.451 & -2.453 & 0.0095  \\
0.25 & 2.427 & -4.493 & -4.479 & 0.0418 &
1.35 & 70.77 & -2.390 & -2.391 & 0.0086  \\
0.30 & 3.495 & -4.335 & -4.332 & 0.0257 &
1.40 & 76.11 & -2.331 & -2.332 & 0.0078  \\
0.35 & 4.757 & -4.196 & -4.194 & 0.0193 &
1.45 & 81.64 & -2.273 & -2.274 & 0.0071  \\
0.40 & 6.213 & -4.070 & -4.064 & 0.0175 &
1.50 & 87.37 & -2.218 & -2.218 & 0.0065  \\
0.45 & 7.863 & -3.953 & -3.943 & 0.0174 &
1.55 & 93.29 & -2.163 & -2.164 & 0.0059  \\
0.50 & 9.708 & -3.842 & -3.830 & 0.0176 &
1.60 & 99.40 & -2.110 & -2.111 & 0.0054  \\
0.55 & 11.75 & -3.734 & -3.721 & 0.0177 &
1.65 & 105.7 & -2.059 & -2.059 & 0.0050  \\
0.60 & 13.98 & -3.628 & -3.617 & 0.0175 &
1.70 & 112.2 & -2.008 & -2.008 & 0.0046  \\
0.65 & 16.41 & -3.525 & -3.516 & 0.0170 &
1.75 & 118.9 & -1.958 & -1.958 & 0.0043  \\
0.70 & 19.03 & -3.423 & -3.417 & 0.0164 &
1.80 & 125.8 & -1.909 & -1.909 & 0.0040  \\
0.75 & 21.84 & -3.325 & -3.322 & 0.0158 &
1.85 & 132.9 & -1.861 & -1.861 & 0.0038  \\
0.80 & 24.85 & -3.229 & -3.229 & 0.0152 &
1.90 & 140.2 & -1.814 & -1.814 & 0.0036  \\
0.85 & 28.05 & -3.137 & -3.138 & 0.0146 &
1.95 & 147.7 & -1.768 & -1.767 & 0.0034  \\
0.90 & 31.45 & -3.048 & -3.051 & 0.0141 &
2.00 & 155.3 & -1.722 & -1.722 & 0.0032  \\
0.95 & 35.04 & -2.962 & -2.966 & 0.0137 &
2.05 & 163.2 & -1.678 & -1.677 & 0.0031  \\
1.00 & 38.83 & -2.880 & -2.884 & 0.0134 &
2.10 & 171.2 & -1.634 & -1.634 & 0.0030  \\
1.05 & 42.81 & -2.800 & -2.805 & 0.0132 &
2.15 & 178.5 & -1.591 & -1.591 & 0.0029  \\
1.10 & 46.98 & -2.724 & -2.729 & 0.0131 &
2.20 & 187.9 & -1.549 & -1.549 & 0.0028 \\
\hline
       \end{tabular}
       \end{footnotesize}
     \end{center}  
   \end{minipage} 
\end{center}
\end{table}
\vspace{-2mm}
 However the numerical value of $\Delta_{0}$ is extremely small, 
 namely less than $2 \times 10^{-3}$ or less than 10\% of the error 
 of $\tilde{K}_{0}^{once}(\nu)$ for $0< \nu < 3$.
   The smallness of $\Delta_{0}$ comes from the facts that  
firstly the meromorphic function reproduce $X_{0}(\nu)$ well in the 
wider domain of $ T_{lab} \stackrel{<}{\sim} 180 MeV.$, secondly in 
the neighborhood of $T_{lab}=248 \; MeV.$, where the phase shift 
passes through zero, the spectral function Im $ h_{0}(\nu')/\nu'$ is 
small and thirdly in higher $\nu'$ region the spectral 
function\cite{Arndt} is suppressed by the factor $\nu'$, 
because we are computing the once sutracted Kantor amplitude.   
The last column of Table 3 is the error of $\tilde{K}_{0}^{once}
(\nu)$, which is defined by 
\begin{equation}  
  \Delta \; \tilde{K}_{0}^{once}(\nu) =\frac{1}{2} 
  \{ |\tilde{K}_{0}^{  [uc ] \; once }(\nu) -\tilde{K}_{0}^{ [sc ] 
  \; once }(\nu)| +|\tilde{K}_{0}^{ [lc] \; once }(\nu)
   -\tilde{K}_{0}^{[sc] \; once }(\nu)| \}  .
\end{equation}

       In figure 4, $-\tilde{K}_{0}^{once}(\nu)$ is plotted against 
 $T_{lab}$, and in which the diamonds are [sc] and the triangles 
  are  [fc] respectively.  Figure 5 is the enlarged graph of 
 the lower energy part of figure 4.   The curves in the figures are 
 that $(L)_3$ is the 3-parameter fit by the spectral function of 
 the long range force, whereas $(sa)_3$ (dotted curve) and 
 $(sb)_3$ (dash curve) are the fits by 
 the short range potential with three free parameters.   Details of 
 the curves will be explained in sections 5. 
   A characteristic feature of $- \tilde{K}_{0}^{once}(\nu)$ is 
   that it has a sharp cusp at $\nu=0$, contrary to what is expected 
 in the meson theory of the nuclear force, where the location of the 
 two-pion exchange spectrum is `far away' compared to the width of 
 the cusp.   In the one-pion exchange term of Eq.(6), 
 $g^2/4 \pi=14.4$ is used as the $\pi$-$N$ coupling 
 constant\cite{GpiN}\cite{GpiN2}.  However modifications to the 
 different values\cite{nij90}\cite{nij93} of  $g^2/4 \pi$ are 
 straightforward if we remember Eq.(4).   
 It must be pointed out that as the coupling constant $g^2/4 \pi$ 
 decreases the cusp of $-\tilde{K}_{0}^{once}(\nu)$ becomes 
 more prominent.
 
 \begin{figure}[htbp]
\begin{minipage}{6.8cm}
\includegraphics[width=.99\textwidth,height=5.5cm]{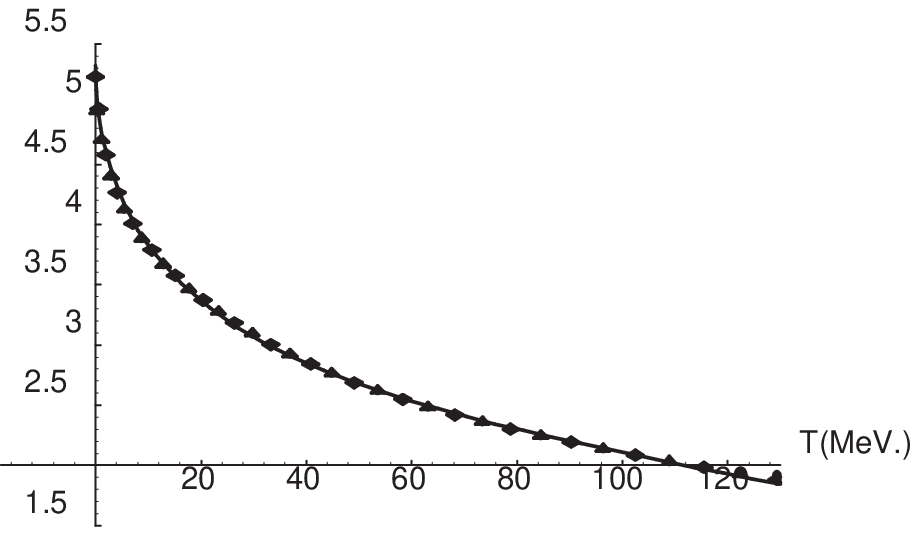}
\caption{{\footnotesize
 $-\tilde{K}_{0}^{once}(\nu)$ is plotted against $T_{lab}$ in 
  $T_{lab} < 125 \; MeV.$.   The diamonds and triangles are  
 $-\tilde{K}_{0}^{once}(\nu)$ calculated from $X_{0}(\nu)$ of 
  the [sc] and the [fc] fits, whose parameters are listed in 
Table 1, respectively.   
 The curve is the fit by the spectrum of the long 
range force in the range $ 0.6\;MeV. < T_{lab} <125\; MeV.$ .  }}
\end{minipage}
\hfill
\begin{minipage}{6.8cm}
\includegraphics[width=.99\textwidth,height=5.cm]{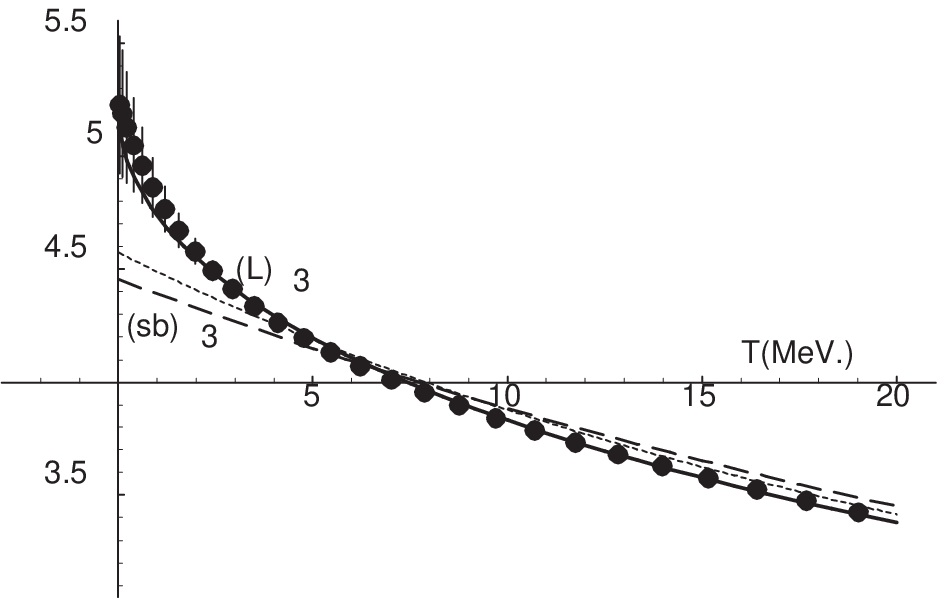}
 \caption{{\small
The enlarged graph of $-\tilde{K}_{0}^{once}(\nu)$  of [sc] is plotted 
 in $ T_{lab} < 20 \; MeV.$.  The three curves are 
3-parameter fits to the data in $ 0.6\;MeV. < T_{lab} <125\;
 MeV.$. 
The curve $(L)_3$ is the fit by the spectrum of the long 
range force, whereas the curves of $(s \; a)_3$ (dots)
 and $(s \; b)_3$ (dash) 
are the fits by the short range potentials. }}
\end{minipage}
\end{figure}

     In the next two sections, $\tilde{K}_{0}^{once}(\nu)$  of 
Table 3 will be analysed to understand the physical meanings of the 
cusp at $\nu=0$. 
In section 4, the cusp will be fitted by the spectrum of the long 
range force, in which the power $\gamma$ of the threshold behavior 
of the spectrum  will be searched.  On the other hand, 
in section 5 it will 
be examined whether the short range nuclear potential 
$V(r)$ can reproduce well the cusp of $\tilde{K}_{0}^{once}(\nu)$.   

 \section{Fits by the Spectrum of the Long Range \\ 
 Interactions }

   Figures 4 and 5 indicate that $-\tilde{K}_{0}^{once}(\nu)$ has a 
cusp  pointed upward at $\nu=0$.    If there exists the spectrum of 
the long range force with the attractive sign, appearance of such 
a cusp in   $-\tilde{K}_{0}^{once}(\nu)$ is a natural consequence.   
In this section, we shall determine the power $\gamma$ and the 
strength of the threshold behavior of the spectral function at 
$\nu=0$.    More conventional fit will be tried in the next 
section, in which the possibility to understand the cusp of 
$-\tilde{K}_{0}^{once}(\nu)$ in terms of the ordinary short range 
potential of the nuclear force will be considered.  The integral 
representation of the scattering amplitude $\tilde{A}(s,t)$ of the 
non-OPE part is
\begin{equation}   
  \tilde{A}(s,t)=\frac{1}{\pi} \int_{t_{0}}^{\infty} dt' 
  \frac{\tilde{A}_{t}(s,t')}{t'-t} \pm  \frac{1}{\pi} 
  \int_{u_{0}}^{\infty} du' \frac{\tilde{A}_{u}(s,u')}{u'-u} \quad ,
\end{equation}
where $t_{0}$ and $u_{0}$ are $4 \mu^2$ for the short range 
potential whereas zero for the long range interaction, respectively.  
 The tilde such as $\tilde{A}$ means the function in which the OPE 
part is already subtracted.   Since we are considering the scattering 
of the identical fermi particles,  the signs between the integrations 
of Eq.(25) are plus for the spin-singlet and minus for the 
spin-triplet states respectively.  Moreover the spectral functions 
of t and u channels are the same, namely $ \tilde{A}_{t}(s,\ast)= 
\tilde{A}_{u}(s,\ast)$. The partial wave projection of Eq.(25) gives
\begin{equation}  
\tilde{h}_{0}(\nu)=\frac{1}{\pi}  \int_{0}^{\infty} dt \tilde{A}_{t}
(s,t) \frac{1}{2 \nu} Q_{0}(1+\frac{t}{2\nu}) + ( t \rightarrow u ) 
\; .
\end{equation}        
Therefore the spectral function of the left hand cut of the 
S-wave amplitude $\tilde{h}_{0}(\nu)$ is
\begin{equation}  
{\rm Im} \tilde{h}_{0}(\nu)= \frac{1}{4 \nu} \{ \int_{t_{0}}^{-4 \nu} 
dt \tilde{A}_{t}(s,t) +\int_{u_{0}}^{-4 \nu} du \tilde{A}_{u}(s,u) \} 
\quad for \quad \nu < -t_{0}/4 \;  .
\end{equation}

    Let us search the value of the power $\gamma$ of the spectral 
function of $ \tilde{A}_{t}(4 m^2, t)$ at $t=0$ or of ${\rm Im} 
\tilde{h}_{0} (\nu)$ at $\nu=0$.  Because of Eq.(27), the powers 
$\gamma$'s of these functions are the same.   Since $ \tilde{A}_{t}
(4 m^2, t)$ directly relates to the potential by the Laplace 
transformation
\begin{equation}  
  r V(r) = - \frac{1}{\pi m^2} \int_{0}^{\infty} dt' 
  A_{t}(4 m^2, t') e^{-r \sqrt{t'}} \quad ,
\end{equation}
we shall assume the functinal form of  $ \tilde{A}_{t}(4 m^2, t)$ 
and search the parameters involved.  In the search of the power 
$\gamma$, three-parameter form of the spectral function will be 
used, which is
\begin{equation}   
   \tilde{A}_{t}(4 m^2, t)= \pi C' t^{\gamma} \exp [-\beta t] \quad ,
\end{equation}
where the exponential factor is necessary to make the integration 
Eq.(26) convergent.\footnote{Moreover the parameter $\beta$ plays the 
roll to incorporate the two-pion exchange spectrum in Eq.(29).} 

     In the search, 68 points of $\tilde{K}_{0}^{once}(\nu)$ in 
$ 0.6 MeV. < T_{lab} < 125 MeV.$ are fitted.  The fitted points are 
chosen to be equispacing with respect to $\sqrt{\nu}$, or more 
precisely $\sqrt{\nu_{n}}=0.025n$ with $5 \leq n \leq 72$.    
The parameters $(\gamma \; , \; \beta \; , \; C' \;)$ and $\chi$ of 
the best fits are
\begin{eqnarray}  
\gamma = 1.543 \; , \quad  \beta=0.06264 \; , \quad C'=0.1762 
\quad and \quad \chi=0.441 & for &\mbox{[sc]}   \\
\gamma = 1.569 \; , \quad  \beta=0.06544 \; , \quad C'=0.1734 
\quad and \quad \chi=0.343 & for &\mbox{[$\,$fc$\,$]}       
\end{eqnarray} 
respectively.   The curves in figures 4 and 5 are the fits of [sc] 
(full line) and [fc] (dash line) respectively.

         In figure 6,  $\chi(\gamma)$ is plotted against $\gamma$, 
in which the full line is [sc] and the dash line is [fc] respectively. 
 $\chi(\gamma)$ is obtained by making the chi-square search for 
fixed $\gamma$.   From the curves, minimum points are determined, 
which are $\gamma_{min}= 1.543 \pm 0.055$ and $1.569 \pm 0.044$ for 
[sc] and [fc] respectively.     
The range of the uncertainty of $\gamma_{min}$ is estimated by the 
condition that $ \chi(\gamma) \leq 1.25 \chi_{min}$, in which 
$\chi_{min}$ is the minimum value of $\chi(\gamma)$.      
It is remarkable that the value of the power $\gamma$  is close to 
what is expected in the case of the van der Waals potential of the 
London type namely to $\gamma=1.50$.

\begin{figure}[htbp]
\begin{minipage}{6.8cm}
\includegraphics[width=.99\textwidth,height=5.0cm]{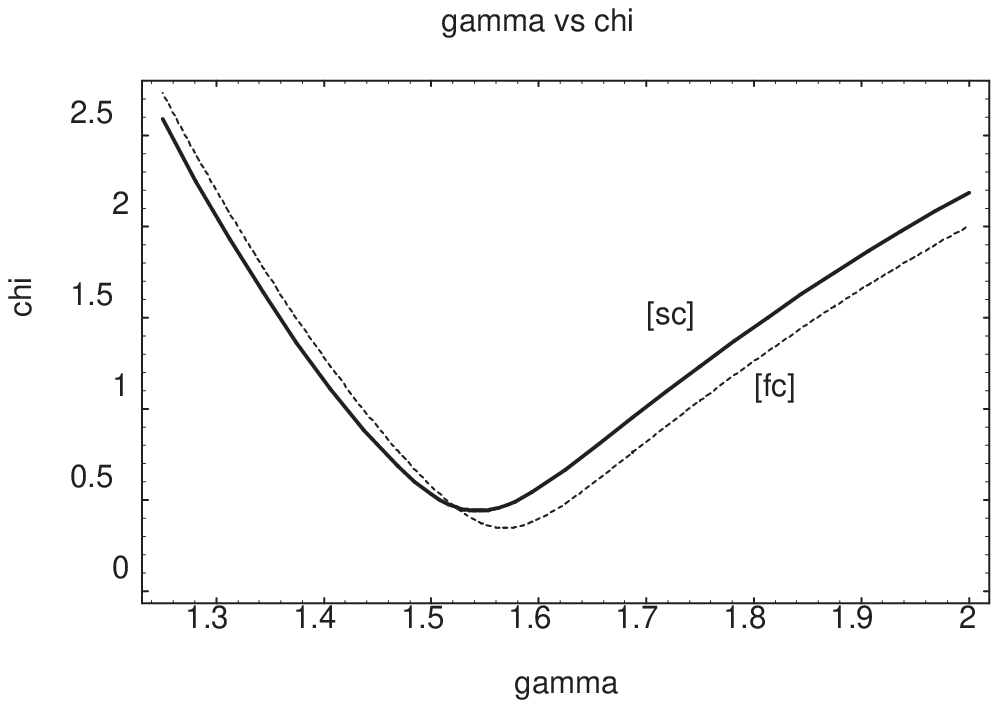}
\caption{{\footnotesize
 $\chi(\gamma)$ are plotted against $\gamma$. The curves are the 
fits to the $-\tilde{K}_{0}^{once}(\nu)$ of [sc] (full line) and 
of [fc] (dash line) respectively. The minimum is close to 
$\gamma=1.5$ of the Van der Waals force of the London type.
  }}
\end{minipage}
\hfill
\begin{minipage}{6.8cm}
\includegraphics[width=.99\textwidth,height=5.0cm]{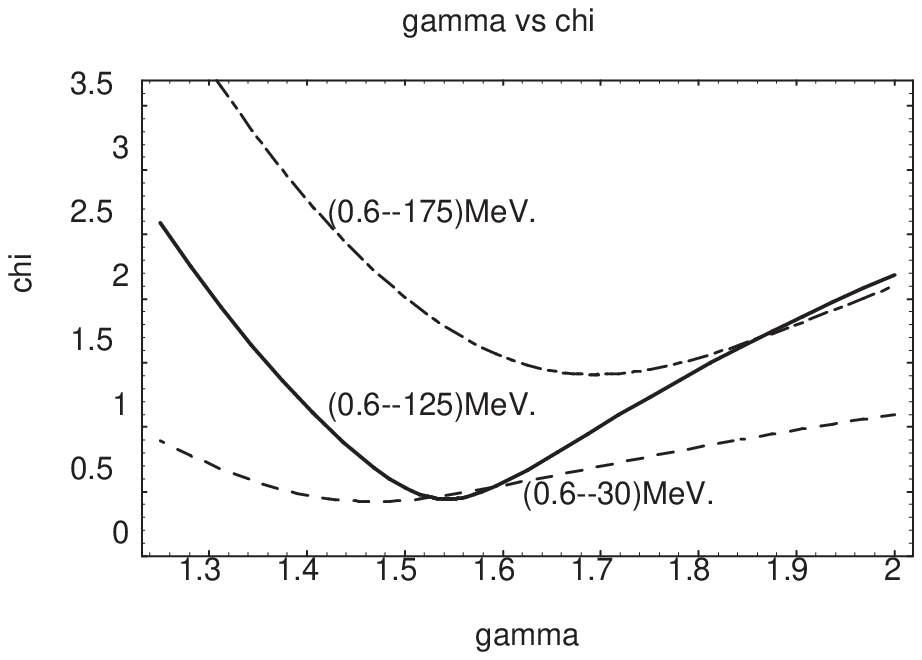}
\caption{{\footnotesize 
$\chi(\gamma)$ are plotted against $\gamma$ for [sc].  
The three curves correspond to the different fitted regions. 
They are $T_{lab}=$ $(0.6 \sim 30) MeV.$ , 
$(0.6 \sim 125) MeV.$ and $(0.6 \sim 175) MeV.$ respectively.}}
\end{minipage}
\end{figure}

     However before we discuss about the physical meaning of the 
value that $\gamma \approx 1.5$, it is necessary to examine the 
dependency of $\gamma_{min}$ on the fitted range of energy. In figure 
7, three curves of $\chi(\gamma)$ with different fitted energy 
domains are shown.    They are $(0.6 \sim 30)$ MeV. and 
$(0.6 \sim 175)$ MeV., in addition to the domain $(0.6 \sim 125)$ 
MeV., which was shown in fig.6.   The figure indicates that as 
the energy domain shrinks, $\gamma_{min}$ moves slowly to samaller 
value.   For $ 0.6 MeV. < T_{lab} < 30 MeV.$ the best fit is 
that $\gamma_{min}=1.467 \pm 0.111$ and $\chi_{min}= 0.423$ for [sc]. 
 On the other hand, as the energy region expands, $\gamma_{min}$ 
increases and moreover the value of $\chi_{min}$ increases rapidly.   
 Therefore the 3-parameter function of Eq.(29) is not suitable for 
the fit to so many data points.   In fact, for the domain $ 0.6 MeV. 
\leq T_{lab} \leq 175 MeV.$, the curve in the figure indicates that 
$\gamma_{min}=1.695 \pm {0.192 \atop 0.150} $ and $\chi_{min}= 1.410$ 
for [sc].   Therefore even if we change the energy domain, the power 
$\gamma$ still remains in the neighborhood of 1.5 . 
 
\section{Fits by the Spectral Function of the Short Range Interactions}

        In the previous section, it was found that the cusp in 
$\tilde{K}_{0}(\nu)$ obtained in section 3 was reproduced well by the 
spectral function $\tilde{A}_{t}(4m^2,t)$ of the long range 
interaction given in Eqs.(29),(30) and (31).  However it is important 
to ask whether the cusp can be understood also in the framework of 
the more conventional approach, namely in terms of the spectral 
function of the short range interaction expected in the meson theory. 
 It is well known that, in the meson theory of the nuclear force, 
 the spectral function of the two-pion exchange is
\begin{equation}    
\tilde{A}_{t}(4m^2,t)= \left\{ \begin{array} {ll}
\qquad  0  & {\rm in} \quad t \leq 4 \mu^2 \\ 
 \pi C' _{2 \pi} (t-4 \mu^2)^{1/2} + \cdots  & {\rm in} 
 \quad t > 4 \mu^2
\end{array}
\right.
\end{equation}
, in the neighborhood of the two-pion threshold at $t=4 \mu^2$.   
 The power 1/2 of Eq.(32) comes from the functional form of the 
 phase volume of the two-particle state exchanged.  The coefficient 
$C'_{2 \pi}$ can be estimated from the phase shifts of the $\pi$-$N$ 
and the $\pi$-$\pi$ scatterings by using the generalized unitarity of 
the dispersion technique\cite{matom}.    From Eqs.(27) and (32), the 
spectral function of the left hand cut of $\tilde{h_{0}}(\nu)$, which 
is the same as that of $\tilde{K_{0}}(\nu)$,  becomes 
\begin{equation}  
 { \rm Im} \tilde{h}_{0} (\nu') = \left\{ \begin{array} {ll}
 \qquad 0  & {\rm in} \quad \nu' >- 1 \\ 
\frac{8 \pi}{3 \nu'}  C'_{2 \pi} (-\nu' - 1)^{3/2} + \cdots  & 
{\rm in} \quad \nu' \leq -1 \quad .
\end{array}
\right.
\end{equation}
 Since the power of the spectral function at the threshold $\nu=-1$ 
increases to 3/2, the distance of the average location of the 
spectrum from $\nu=0$ is much larger compared to that of the 
threshold point.   With such a spectral function, it will be not easy 
to reproduce the cusp of $\tilde{K}_{0}^{once}(\nu)$ observed in 
$ \nu \ll 1$, where in our unit $\nu=1$ corresponds to 
$ T_{lab}=38.83 MeV.$ .     On the other hand, the updated nuclear 
potentials are believed to fit well to the precise phase shift 
data.\cite{Parispot}\cite{Bonnpot}\cite{NIJM94}\cite{ARGONNE}.  

    In this section, we shall examine the updated nuclear potentials 
more closely, and in which we shall concentrate on the cusp at 
$\nu=0$.  Among various potentials, we shall choose the regularized 
Reid potential updated by the Nijmegen group (Reid 93) as an 
example\cite{NIJM94}, in which 50 phenomenological parameters are 
searched to fit to 1787 p-p data and 2514 n-p data in the energy 
range of $ 0 < T_{lab} < 350 MeV.$.   It is remarkable that by using 
such potential they achieved excellent fits $\chi^2=1.03$ per datum, 
which is close to the chi-square value $0.99$ of the energy dependent 
phase shift analysis.  In the Reid93, the S-wave potential of the 
non-OPE part of the p-p scattering involves five parameters: 
\begin{equation}   
V_{pp}(^{1} S_{0}) =A_{2}Y(2)+A_{3}Y(3)+A_{4}Y(4)+A_{5}Y(5)+A_{6}Y(6)
 \quad ,
\end{equation}
where the regularized Yukawa functions of the range $1/(p \mu )$ are
\begin{equation}   
Y(p)=p \frac{1}{r} [ e^{-p \bar{\mu} r} - e^{- \Lambda r}
(1+\frac{\Lambda^2- p^2 \bar{\mu}^2}{ 2 \Lambda^2} \Lambda r ) ]
\end{equation}
and $\bar{\mu}$ is the average masses of the three types of pions 
and $\Lambda=8 \bar{\mu}$.    Contrary to the original Reid 
potential\cite{Reid}, this potential has a characteristic feature 
that the $Y(2)$ term exists.  Such term gives rise to a pole at 
$t=4 \mu^{2}$ in the scattering amplitude $A(s,t)$, whereas in the 
meson theory $t=4 \mu^{2}$ is merely the starting point of the 
continuous spectrum of the two-pion exchange.   Therefore it is not 
desirable from the point of view of the meson theory to have the 
$Y(2)$ term in Eq.(34).  On the other hand the inclusion of the 
$Y(2)$ term is necessary, in fitting the very precise low energy data 
of the p-p scattering, to reduce the the chi-square value.   
The contradictory situation already indicates the difficulty to 
reproduce the cusp from the far-away spectral function of the meson 
theory of the nuclear force.   

    In figure 5, results of the 3-parameter fits to the data points 
in $ 0.6\;MeV. < T_{lab} <125\; MeV.$ are compared.  
The curve $(L)_3$ is the fit by the spectrum of the long range force, 
while the curves of $(s \; a)_3$ and $(s \; b)_3$ are the fits by 
the short range potentials of the Reid types of Eq.(34).  
The $(s \; a)_3$ is the sum of $Y(2)$, $Y(3)$ and $Y(4)$, on the 
other hand $(s \; b)_3$ is the sum of $Y(3)$, $Y(4)$ and $Y(5)$, 
where $Y(p)$ is the regularized Yukawa potential of range $1/p$ 
defined in Eq.(35).   The $\chi$-values per datum of $(L)_3$, 
$(sa)_3$ and $(sb)_3$ are 0.441, 1.82 and 3.11 respectively.    
It is evident that the short range potentials with three parameters 
cannot reproduce the cusp at $\nu=0$.  It is interesting that the 
inclusion the $Y(2)$ term always reduce the $\chi$-value, this is 
because the lack of the spectrum of small $t$ is the reason of the 
deviation from the data points of  $ \tilde{K}_{0}^{once}(\nu) $.    

\begin{figure}[htbp]
 \includegraphics[width=.9\textwidth,height=7.0cm]{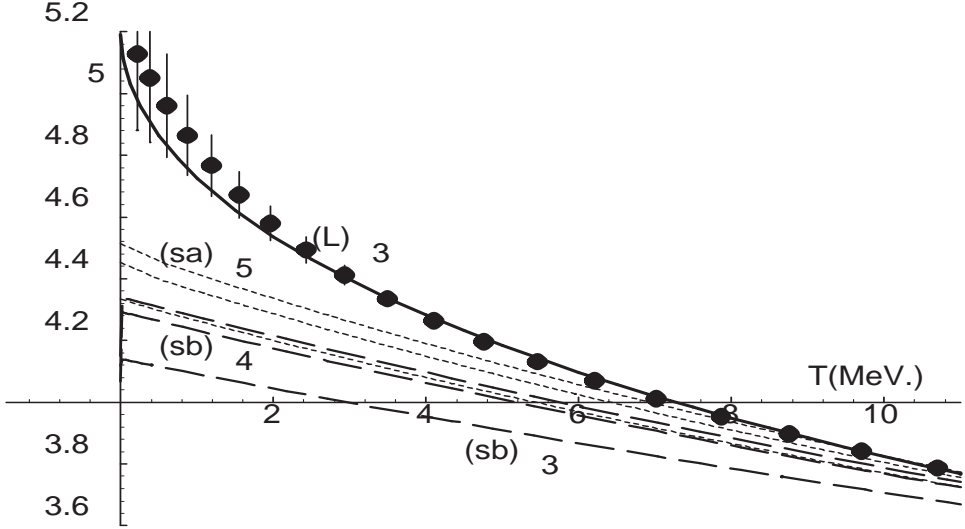}
\caption{{\footnotesize
The seven curves are the fits to $\; -\tilde{K}_{0}^{once}(\nu)\; $ of 
[sc] in the off-cusp region $T_{lab}=$ $(20 \sim 125)MeV.$.   
The fitted curves are then extrapolated to $T_{lab}=0$.  
The $(L)_3$ curve is the 3-parameter fit by the spectrum of the long 
range force, on the other hand $(s \; a)_n$ (dotted curves) and 
$(s \; b)_n$ (dash curves) are the n-paremeter fits by the 
Reid type potentials.   The differences of $(s \; a)_n$ and 
$(s \; b)_n$ are that, although $(s \; a)_n$ involves $Y(2)$ term, 
in $(s \; b)_n$ the Yukawa term of the longest range is $Y(3)$.   }}
\end{figure}

     In figure 8, the seven curves are the fits to 
$\; -\tilde{K}_{0}^{once}(\nu)\; $ of [sc] in the off-cusp 
region $T_{lab}=$ $(20 \sim 125)MeV.$.   
The fitted curves are then extrapolated to $T_{lab}=0$.  
The $(L)_3$ (full line) curve is the 3-parameter fit by the 
spectrum of the long range force of Eq.(29) as before, on the other 
hand $(s \; a)_n$ (dotted line) and $(s \; b)_n$ (dash line) are the
 n-paremeter fits by the Reid type potentials.   
 The differences of $(s \; a)_n$ and $(s \; b)_n$ are 
that, although $(s \; a)_n$ involves $Y(2)$ term, in $(s \; b)_n$ the 
Yukawa term of the longest range is $Y(3)$.   In general as the 
numbers of the parameters $n$ increase, the curves come closer 
to the data points.  However the curves of Fig.8 indicate that 
the curves of the short range interaction $(s \; b)_n$ cannot 
reproduce the cusp of the data points at $\nu=0$.
 
    In the meson theory, the nearest singularity of 
  $\; -\tilde{K}_{0}^{once}(\nu)\; $ occurs at $\nu=-1$ or in terms 
  of $T_{lab}$ at $-38.83\; MeV.$, which is the threshold of the 
  two-pion exchange spectrum.    
  Since Eq.(33) indicates that the power of of the threshold behavior 
  of the two-pion exchange spectrum at $\nu=-1$ is 3/2, the distance 
  to the average location of the spectrum from the origin is much 
  larger than one.  What is important is that the length of our 
  extrapolation ($\sim$ 0.5) is short compared not only to the size 
  of the fitted domain ($\sim 3.0$) but also to the distance to the 
  left hand spectrum ($ > 1$).   Therefore if the analytic structure 
  expected in the meson theory is correct, the extrapolated curve 
  must approximately follow the data points.  However figure 8 
  indicates that this is not the case.   
  On the other hand, the extrapolated curve of the long range force 
  $(L)_3$ traces the data point nicely.  Therefore we conclude that 
  the analytic structure of the amplitude of the  nucleon-nucleon 
  scattering is different from what is expected in the meson theory 
  of the nuclear force, rather the nuclear force involves the large 
  component of the long range force.   
  
\section{Strong van der Waals Interaction as a \\
Candidate of the Nuclear Force  }

    In section 2, we introduced the S-wave Kantor amplitude of the 
proton-proton scattering $K_{0}(\nu)$, which was free from the 
singularities of the vacuum polarization and the Coulombic 
interaction as well as the unitarity cut.  By subtracting the 
one-pion exchange contribution from $K_{0}(\nu)$, the domain of 
analyticity expands further, and the nearest singularity of the 
non-OPE Kantor amplitude $\tilde{K}_{0}(\nu)$ occurs at $\nu=-1$, 
and which is the branch point of the two-pion exchange spectrum. 
By using the phase shift data of the Nijmegen group\cite{nij88}
\cite{nij90}, the once subtracted non-OPE Kantor amplitude 
$\tilde{K}_{0}^{once}(\nu)$ defined in Eqs.(6)and (18) was 
calculated in section 3.   The appearance of the cusp at $\nu=0$ 
is rather 
surprising, because in the meson theory of the nuclear force, 
$\tilde{K}_{0}^{once}(\nu)$ is regular at least in $|\nu|<1 $. 
    
    It turns out that the conventional fits of the spectra of the 
short range forces cannot reproduce the cusp properly even if the 
potentials with five parameters given in Eq.(34) are used in the 
fitting.   On the other hand, if we use the spectra 
of the long range force, three parameters of Eq.(29) are sufficient 
to reproduce the cusp of $\tilde{K}_{0}^{once}(\nu)$.   
In summary, the short range condition, that Im$h_{0}(\nu) \equiv 0$ 
in $\nu>-1$, is too severe to reproduce the cusp in $|\nu| \ll 1$.      
Therefore it is natural to regard the cusp as an effect of the long 
range interaction between hadrons.  The power $\gamma$ of the 
threshold behavior was searched in section 4, and it turned out that 
$\gamma=1.543 \pm 0.056$ from the multi-energy data [sc] and 
$\gamma=1.569 \pm 0.046$ determined from the fit [fc], respectively.  
Since the power $\alpha$ of the asymptotic behavior of the long range 
potential $V(r)$ is related to the power $\gamma$ by 
$\alpha=2 \gamma +3$, the values of $\gamma$ imply 
$\alpha=6.09 \pm 0.11$ and $\alpha=6.17 \pm 0.09$ for [sc] 
and [fc] respectively.

     It is remarkable that the values of $\alpha$ are close to that 
of the van der Waals potential of the London type ( $\alpha=6$ ).  
Although  $\alpha$ can assume any value from $-1$ (Coulomb) to 
infinity (Yukawa), Nature chooses a special value in the neighborhood 
of $\alpha=6$.    Therefore it is difficult to believe the occurrence 
of the value $\alpha = 5.98 \sim 6.26$ is merely accidental, rather 
the ordinary mechanism to produce the van der Waals interaction must 
be working.    It is well-known in the atomic and molecular physics 
that, when the underlying Coulombic force exists, the dipole-dipole 
interaction occurs between the neutral composite particles, and the 
two-step process of such interactions causes the attractive 
potential\cite{paulingw}.   The form of the potential between 
particle 1 and particle 2 is
\begin{equation}   
 V(R)=- \mathop{{\sum}'}_{n}  \frac{(H')_{0 n} 
 (H')_{n 0}}{E_{n} - E_{0}} \quad ,
\end{equation}
where $H'$ is the hamiltonian of the dipole-dipole interaction 
\begin{equation}  
 H'=\frac{ ^{\ast} e ^2}{R^3} (x_{1}x_{2} + y_{1}y_{2} -2 z_{1}z_{2})
\quad ,
\end{equation}
in which $^{\ast} e^2 $ is the strength of the underlying Coulombic 
interaction.  

       It is important to notice that each term in the summation of 
Eq.(36) is positive definite.   Therefore if we replace all the 
denominators $ (E_{n} - E_{0})$ in Eq.(36) by the first excitation 
energy $(\Delta E_{1})$, then by using the closure property the 
upper bound of the strength $C_{6}$ of the van der Waals potential, 
which is introduced by $ V(R)=-C_{6}/R^6 + \cdots $, is obtained.   
It is\cite{paulingw}
\begin{equation}  
 C_{6} < \frac{R^6 ({H'} ^2)_{00}}{(\Delta E_{1})} = \frac{2}{3} 
 \frac{ ^{\ast}e^4 }{(\Delta E_{1})} \; \overline{r_{1}^2} \; \; 
 \overline{r_{2}^2} \; \; .
\end{equation}
 On the other hand, if we retain only the terms of the first excited 
 states in the summation, Eq.(36) gives the lower bound of $C_{6}$, 
 namely
\begin{equation}  
 C_{6} >  \frac{2}{3} \frac{ ^{\ast}e^4 }{(\Delta E_{1})} 
 \tilde{R_{1}}^2 \; \tilde{R_{2}}^2 \quad ,
\end{equation}
where $\tilde{R_{i}}^2$ relates to the dipole transition amplitude 
 $\tilde{z}$ by
\begin{equation}  
  \tilde{R}^2= 3 |\tilde{z}|^2 \qquad {\rm with} \quad \tilde{z}= 
\int d^3 r \psi^{\ast}_{1,1,0} \; (r \cos \theta) \; \psi_{0,0,0} \; .
\end{equation}
 Since we know the value of $C_{6}$, Eqs.(38) and (39) will be used 
in turn to give the lower and the upper bounds of the strength of 
the underlying Coulombic force $^{\ast} e^2$ respectively, 
and which are
\begin{equation}   
 \sqrt{\frac{3}{2} C_{6} (\Delta E_{1})}\quad \frac{1}
 {\overline{r^2}} < \; ^{\ast} \! e^2  <  \sqrt{\frac{3}{2} 
 C_{6} (\Delta E_{1})} \quad \frac{1}{\tilde{R}^2} \; 
\end{equation}
 
     The numerical values of $C'$ and $\beta$ of Eq.(29) for 
$\gamma=1.5$ are obtained by shrinking the energy range to be fitted. 
The results are
\begin{eqnarray}
\gamma & =1.5 & \; , \quad \beta=0.0547 \quad and \quad  C'=0.174 
\qquad for \quad  [sc] \\
\gamma & =1.5 & \; , \quad \beta=0.0541 \quad and \quad  C'=0.172 
\qquad for \quad  [\; fc \;] 
\end{eqnarray}
in the unit of the neutral pion mass, and the energy range of the 
fits are  $(0.6 \sim 34)$ MeV. and $(0.6 \sim 39)$ MeV. for [sc] 
and [ fc ] respectively.   Since the coefficient $C_{\alpha}$ of 
the asymptotic potential is
\begin{equation}   
 C_{\alpha} =C' \frac{2 \Gamma (\alpha -1)}{m^2}  \qquad {\rm with} 
 \quad \alpha=2 \gamma +3 \quad ,
\end{equation}
$C_{6}= 0.9933 C'$ in our unit.     If we use the mass difference 
of $\Delta(1232)$ and $N$ as the first excitation energy, namely 
$\Delta E=2.18$, then the lower bound of $^{\ast}e^2$ becomes
\begin{equation}  
 ^{\ast} e^2 >  \frac{\sqrt{3.27 C'}}{\overline{r_{1}^2}} .
\end{equation}

     To go further we need the size of the composite particle, the 
radius of proton in particular.   Although we do not have much 
information on the radius, we shall consider two cases.  In the 
first case, $\overline{r_{1}^2}$ is $ ( 0.7 {\rm fm.})^2$, which 
comes from the nucleon form factor.    In another case 
$\overline{r_{1}^2}$ is set equal to $\beta$ of Eqs.(42) and (43), 
because at the range  $r=\sqrt{\beta}$ the potential starts to 
deviate appreciablly from the asymptotic form of the van der Waals 
potential of $V(R) \sim - C_{6}/ R^6$ and therefore $\sqrt{\beta}$ 
may be regarded as the size of the composite particle.   
The results are :
\begin{eqnarray*}
  ^{\ast}e^2 > 3.27  \qquad & (&{\rm  from \; the \; form 
  \; factor}\; ) \\
  ^{\ast}e^2 > 13.8  \qquad & (&{\rm  from} \quad 
  \overline{r_{1}^2}=\beta \; ) \; .
\end{eqnarray*}
As it is expected, the underlying Coulombic force is super-strong, 
because the induced van der Waals force has already the order of 
magnitude of the strong interaction.  

   The coupling constants $^{\ast}e^2$ of the QED and the QCD are 
too small, they are $(137.036)^{-1}$ and 0.32 respectively.    
On the other hand, the Coulombic force of the magnetic monopoles 
is an imortant candidate of the underlying Coulombic interaction.   
It is well-known from the charge quantization condition of Dirac 
that\cite{monopole} 
\begin{equation}  
 ^{\ast} e^2 = \frac{137.036}{4} .
\end{equation}
 Therefore $^{\ast} e^2$ of the magnetic charge is large enough to 
 satisfies the condition of the lower bound, moreover it is not very 
 large compared to the value of the lower bound.   This situation is 
 satisfactory, because the upper bound of Eq.(41) must have the value 
 of the same order of magnitude as the lower bound.
 
     Since the Coulombic interaction between the magnetic monopoles 
is a suitable candedite of the underlying dynamics of the nuclear 
force, it is worthwhile to repeat briefly the magnetic monopole model 
of hadron\cite{dyon}.  Although it is overshadowed by the QCD, it 
still has virtue to be considered.
  The monopole model of hadron is essentially the quark model, in 
which the quarks bear the magnetic charges.  Such a fundamental 
particle is often called dyon, since it is doublly charged, namely 
electrically and magnetically charged.  Because of the superstrong 
Coulombic force between magnetic charges, the dyons form the bound 
states of magnetic charge zero, and such composite particles are 
identified with the hadrons. 
        In a sense, hadrons are regarded as `` magnetic atoms '' 
in this model\cite{matom}.    Therefore it is not surprising to 
observe the strong van der Waals forces between hadrons.    
In this article we actually observed the strong van der Waals 
interaction in the proton-proton scattering.   Since the van der 
Waals interaction is universal\cite{universal}, we may expect to 
observe such a force also in other scattering processes.  However 
because of the lower precision of the data, the elimination 
of the two-pion exchange cut is necessary at least in the threshold 
region of the spectrum.   After carrying out such eliminations, we 
can actually  observe the long range force in the P-wave amplitudes of 
the $\pi$-$N$ and the $\pi$-$\pi$ scatterings\cite{paiN}\cite{paipai}.            
\section{Remarks and Comments }
     
    In this paper we searched for the long range force in the S-wave 
amplitude of the proton-proton scattering.  In order to see the 
extra singularity of the long range force in $h_{0}(\nu)$ at $\nu=0$ 
as clearly as possible, we constructed a regular function $K_{0}(\nu)$ 
from the modified effective range function $X_{0}(\nu)$ of the 
proton-proton scattering, where the function $K_{0}(\nu)$ does not 
have the normal singularities in the neighborhood of $\nu=0$.   
By separating the OPE part and then making the once subtracted 
function $\tilde{K}_{0}^{once}(\nu)$, it becomes much easier for us 
to observe the extra singularity at $\nu=0$ when it exists.  
It is impressive to observe that $\tilde{K}_{0}^{once}(\nu)$ has a 
sharp cusp at $\nu=0$, and the form of which is close to the square 
root type, namely to $\tilde{K}_{0}^{once}(\nu)=(C''_{0} + C'' 
\sqrt{\nu})$ with positive $C''$.   The $\sqrt{\nu}$ singularity 
is exactly what is expected in the van der Waals interaction of the 
London type.  Moreover the observed sign of $C''$ indicates that the 
long range  force is attractive.

    If we state it more precisely, when the spectral function has 
the form of $A_{t}(4 m^2,t) = \pi C' t^{\gamma} \exp(-\beta t)$,  
the chi-square fit to the multi-energy phase shift data\cite{nij88}
\cite{nij90} determines the three parameters involved.    
They are  $\gamma =1.54$, $\beta=0.063 $ and $C'=0.175$ in the unit 
of the neutral pion mass.   It is interesting to see that the 
spectral function $A_{t}( 4 m^2,t)$ has a peak at $\sqrt{t}=4.9$ 
namely at 660MeV..   On the other hand, the location of the peak of 
the spectrum $A_{t}(4 m^2,t)/t $ of the once subtracted amplitude is 
$\sqrt{t}=2.93$ or 395MeV..   When we do not need the very precise 
description of the low energy phase shifts, the spectral function can 
be replaced by a $\delta$-function located somewhere between 
400 and 650MeV..   Such a $\delta$-function may 
correspond to the fictitious $\sigma$-meson of the one-boson 
exchange model of the nuclear potential\cite{OBE2}\cite{paipaid}
\cite{sigma}.  

       Since the power $\gamma$ of the threshold behavior of the 
spectral function $A_{t}(4 m^2,t)$ is close to that of the van der 
Waals force of the London type, we may assume that the ordinary 
mechanism, which produce the van der Waals force, is working.   
Then the strength $^{\ast}e^2$ of the underlying Coulombic force can 
be estimated.   This is possible because there are the well-known 
inequalities on the coefficient $C_{6}$ of the van der Waals 
potential\cite{paulingw}, in particular the upper bound of which 
is obtained by using the strength $^{\ast} e^2$ and the radius 
$(\; \overline{ r^2} \;)^{1/2}$ of the composite particle.
     Since we know the strength $C_{6}$ of the van der Waals 
 potential from the fitting, the inequality gives instead the 
lower bound of $^{\ast} e^2$, which depends on size of the 
composite state.    For $(\; \overline{  r^2} \;)^{1/2}=0.5$, 
which is obtained by the measurement of the nucleon form factor, 
we obtain $^{\ast} e^2 > 3.3$, whereas for $( \; \overline{ r^2} 
\; )^{1/2}=\sqrt{\beta}= 0.25$, we obtain $^{\ast} e^2 >14$.   
In any case, the underlying Coulombic force is super-strong.   
Among the interactions of the Coulombic form, the force between 
the magnetic monopoles satisfies this inequalty, because 
$^{\ast} e^2=137.036/4 \; $ due to the charge quantization 
condition of Dirac\cite{monopole}.  Therefore the dyon model of 
hadron, in which the constituent particle dyon bears the magnetic 
charge as well as the electric charge, must be an important 
candidate of the model of hadron\cite{dyon}. 
 
   Finally, in order to confirm the long range force in the 
proton-proton scattering, it is desirable to observe directly the 
difference of the interference patterns by measuring precisely the 
angular distributions of the cross section in the low energy 
experiments.   Our proposal is to observe the interference between 
the repulsive Coulomb force and the attractive long range interaction 
obtained in the present article.    Since the interference pattern of 
the conventional short range foece and the Coulomb force is known, 
the difference can be predicted by using the values of parameters 
$\gamma$, $\beta$ and $C$.   It turns out that the most favorable 
energy is in $T_{lab}= 25MeV. \sim 40MeV.$.   Since the relative 
deviation, 
namely $ \Delta (d\sigma/ d \Omega)\cdot ( d \sigma/d \Omega)^{-1}$, 
has a dip around $ \theta_{c.m.}=10^{\circ}$ with the depth 
$\sim 0.3 \%$, there exist two challenging requirements in the 
measurements.  They are the accuracy and the domain of 
$\theta_{c.m.}$.   However since the shape of the interference 
pattern is important, there is no severe conditions on the 
normalization of the cross section.  Details will be published 
in a separate paper.

\end{document}